 \documentclass[prb,twocolumn,aps,showpacs]{revtex4}

\usepackage{graphicx}
\usepackage{color}
\usepackage{mathrsfs,amsmath,amssymb}

\begin{document}

\title{Three Dimensional Dirac Semimetal and Quantum Transports in
  Cd$_3$As$_2$}

\author{Zhijun Wang, Hongming Weng}

\email{hmweng@aphy.iphy.ac.cn}

\author{Quansheng Wu}

\author{Xi Dai}


\author{Zhong Fang}

\email{zfang@aphy.iphy.ac.cn}

\affiliation{Beijing National Laboratory for Condensed Matter
  Physics, and Institute of Physics, Chinese Academy of Sciences,
  Beijing 100190, China;}

\date{\today}

\begin{abstract}
  Based on the first-principles calculations, we recover the silent
  topological nature of Cd$_3$As$_2$, a well known semiconductor with
  high carrier mobility. We find that it is a symmetry-protected
  topological semimetal with a single pair of three-dimensional (3D)
  Dirac points in the bulk and non-trivial Fermi arcs on the
  surfaces. It can be driven into a topological insulator and a Weyl
  semi-metal state by symmetry breaking, or into a quantum spin Hall
  insulator with gap more than 100meV by reducing dimensionality. We
  propose that the 3D Dirac cones in the bulk of Cd$_3$As$_2$ can
  support sizable linear quantum magnetoresistance even up to room
  temperature.
\end{abstract}


\maketitle

\section{introduction} \label{introduction}

Weyl semimetal is a new topological state of three-dimensional (3D)
quantum matters~\cite{weyl,volovik,monopole,wan, HgCrSe,balents},
different from the 3D topological insulators (TI)~\cite{TIreview,
  TIreview-2,Bi2Se3}.  It can be characterized by Weyl nodes (at Fermi
level) in the bulk and Fermi arcs on the
surfaces~\cite{wan,HgCrSe}. Around the Weyl nodes, the low energy
physics is given as 3D two-component Weyl
fermions~\cite{weyl,volovik}, $H=v\vec{\sigma} \cdot \vec{k}$ (where
$\vec{\sigma}$ is Pauli matrix and $\vec{k}$ is crystal moment), which
carries chirality, left- or right-handed defined by the sign of
velocity $v$. Weyl nodes are stable topological objects as long as
$\vec{k}$ is well defined, and can be viewed as effective magnetic
monopoles in the 3D momentum space~\cite{monopole}.  To get Weyl
semimetal, either time-reversal (TR) or inversion symmetry needs to be
broken~\cite{balents}. Otherwise, there will be double degeneracy for
all $\vec{k}$.  In the case with both TR and inversion symmetries,
however, we may expect a 3D Dirac semimetal state described as
four-component Dirac
fermions, $H=\left( \begin{array}{cc} v\vec{\sigma}\cdot\vec{k} & 0 \\
    0 & -v\vec{\sigma}\cdot\vec{k} \end{array} \right)$, which can be
viewed as two copies of distinct Weyl fermions. Unfortunately, this
expectation is generally not true, because two Weyl nodes with
opposite chirality may annihilate each other if they overlap in
momentum space, and open up a gap in general. Therefore, additional
symmetry is required to protect the 3D Dirac
semimetal~\cite{BiO2,manes,Na3Bi} state and to prohibit the possible
mass term, unless it is at the phase boundary between TI and normal
insulators~\cite{murakami}, a subtle situation hard to be controlled.

The symmetry protected 3D Dirac semimetal has been
discussed~\cite{BiO2,manes} for systems with spin-orbit coupling
(SOC), focusing on special $\vec{k}$ points with
four-dimensional-irreducible-representation (FDIR)~\cite{BiO2}, which
usually appears at the Brillouin Zone (BZ) boundary with
non-symmorphic double space groups. In general, this FDIR requirement
is too strong, and we may expect much wider compound choices by
considering two doubly-degenerate bands with distinct 2D
representations and unavoidable band crossing (protected by
crystalline symmetry). In such case, we may get 3D Dirac points along
the high-symmetry lines rather than high symmetry points at the BZ
boundary. This scenario of Dirac semimetal has been suggested in our
earlier studies on Na$_3$Bi~\cite{Na3Bi}, which is unfortunately not
stable in air. In this paper, we show that a well known compound
Cd$_3$As$_2$ is a symmetry-protected 3D Dirac semimetal with a single
pair of Dirac points in the bulk and non-trivial Fermi arcs on the
surface. It can be driven into a topological insulator, a Weyl
semi-metal, or a quantum spin Hall (QSH) insulator with gap more than
100meV. It can also support sizable linear quantum magnetoresistance
(MR) even up to room temperature. The nice aspect of Cd$_3$As$_2$ is
the high carrier mobility up to 1.5 m$^2$V$^{-1}$s$^{-1}$ at room
temperature and 8.0 m$^2$V$^{-1}$s$^{-1}$ at 4 K, reported about 50
years ago~\cite{mobility}. This makes it a promising candidate for
future transport studies. We will start from the structure and methods
in Sec.~\ref{Methodology}, present the main results in
Sec.~\ref{result}, and finally conclude in Sec.~\ref{Conclusion}.


\section{Crystal Structure and METHODOLOGY} \label{Methodology}

Among the II$_3$-V$_2$-types narrow gap semiconductors, Cd$_3$As$_2$
has drawn crucial attention, because it was believed to have inverted
band structure,~\cite{invert_exp, invert_theory} whereas all others
Cd$_3$P$_2$, Zn$_3$As$_2$ and Zn$_3$P$_2$ have normal band
ordering. In contrast to other inverted band compounds (like HgTe,
HgSe, and $\alpha$-Sn), Cd$_3$As$_2$ belongs to tetragonal symmetry,
and is the representative of this group, which has the splitted
valence band top at $\vec{k}$=0.  The crystal structure of
Cd$_3$As$_2$ is complicated, and can be related to
tetragonally-distorted anti-fluorite structure with 1/4 Cd site
vacancy.  If the distribution of these vacancies is random, one may
treat it by virtual crystal approximation (VCA) for
simplicity~\cite{vcafluorite,vcazb}. However, those vacancies are in
fact ordered even at room temperature, leading to a tetragonal
structure with $D_{4h}^{15}$ ($P4_{2}/nmc$) symmetry (40 atoms per
unit cell, called Structure I hereafter), or a body centered
tetragonal structure with $C_{4v}^{12}$ ($I4_{1}cd$) symmetry (80
atoms per unit cell, called Structure II hereafter), with the later
structure more favored~\cite{Cd3As2Struct}.  This vacancy ordering and
very large cell of Cd$_3$As$_2$ cause historically serious problems
for theoretical studies, and there is no existing first-principles
calculations up to now. We report here the first band structure
calculations of Cd$_3$As$_2$ with its true structures and with SOC
included.

We perform the first-principles band-structure calculations within the
density functional formalism as implemented in VASP~\cite{vasp}, and
use the all-electron projector augmented wave (PAW)~\cite{paw} basis
sets with the generalized gradient approximation (GGA) of Perdew,
Burke and Ernzerhof (PBE)~\cite{pbe} for the exchange correlation
potential. The Hamiltonian contains the scalar relativistic
corrections, and the spin-orbit coupling is taken into account by the
second variation method~\cite{vsoc}. The cutoff energy for the plane
wave expansion was 500 eV and a {\bf k}-point mesh of
$10\times10\times6$ and $6\times6\times6$ are used for the bulk
calculations of Structure I and II, respectively.

For the convenience of our later discussions for the effective low
energy physics, here we briefly introduce our modified second-order
8-band Kane model~\cite{ap_kane} for typical semiconductors. We start
from the standard 4-band second-order Kane model~\cite{ap_kane} for
the case of without SOC, and then introduce additional terms to take
into account the particular tetragonal symmetry of Cd$_3$As$_2$. In
the $k\cdot p$ approximation, considering the low energy $|s \rangle
$, $|p_x\rangle$, $|p_y\rangle$, $|p_z\rangle$ states (as basis)
around $\Gamma$, the modified 4-band Kane model is given as,

\begin{widetext}
\begin{eqnarray*}
  H_4(\vec{k}) & = &\left(\begin{array}{cccc}
      A'{\vec{k}}^{2}+E_s & ik_{x}P &ik_{y}P & ik_{z}P+d\\
      -ik_{x}P & Lk_{x}^{2}+M(k_{y}^{2}+k_{z}^{2})+E_p & Nk_{x}k_{y} &  Nk_{x}k_{z} \\
      -ik_{y}P & Nk_{x}k_{y} & Lk_{y}^{2}+M(k_{x}^{2}+k_{z}^{2})+E_p& Nk_{y}k_{z}\\
      -ik_{z}P+d & Nk_{x}k_{z} & Nk_{y}k_{z} & Lk_{z}^{2}+M(k_{x}^{2}+k_{y}^{2})+E_p-\delta
    \end{array}\right)
\end{eqnarray*}
\end{widetext}
comparing with the strandard isotropic 4-band Kane
model~\cite{ap_kane}, here we consider the anisotropic tetragonal
symmetry, and introduce the paramter $\delta$ for the crystal-field
splitting of $|p\rangle$ orbitals. The other parameter $d$ is
introduced to describe the breaking of inversion symmetry for
Structure II, and it should be zero for Structure I.  Then our
modified 8-band Kane model can be obtained by adding the SOC term as,
\begin{eqnarray*}
H_8(\vec{k})&=&  {\bf{I}} \otimes H_4(\vec{k}) + H_{so},  \\  \\
  H_{so}& = &\frac{\Delta}{2}
  \left(\begin{array}{cccccccc}
  0&0&0&0&0&0&0&0\\
  0&0&-i&0&0&0&0&1\\
  0&i&0&0&0&0&0&-i\\
  0&0&0&0&0&-1&i&0\\
  0&0&0&0&0&0&0&0\\
  0&0&0&-1&0&0&i&0\\
  0&0&0&-i&0&-i&0&0\\
  0&1&i&0&0&0&0&0
\end{array}\right)
\end{eqnarray*}
where \textbf{I} is the $2\times2$ identity matrix and
$\Delta$ denotes the strength of SOC.

\section{Results and discussions} \label{result}

\subsection{Electronic structures and band inversion} \label{bandstruct}

\begin{figure}[tbp]
\includegraphics[clip,scale=0.4]{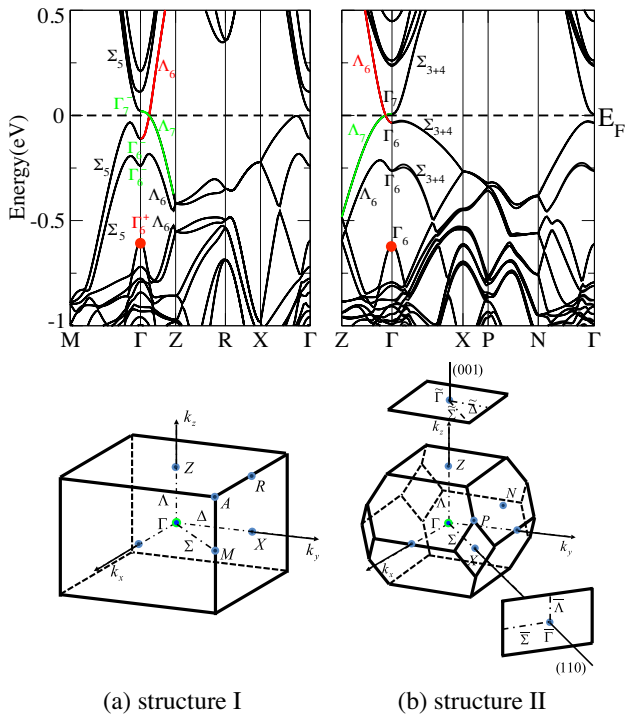}
\caption{\label{ldaband}(Color online) The calculated band structures for
  Cd$_3$As$_2$ with crystal structure I (a) and II (b), as well as the
  definition of high symmetric points in each Brillouin zone (BZ).
  The projected surface BZ for structure II is also shown. The
  representation of selected bands at $\Gamma$ and along high
  symmetric $\vec{k}$ path are indicated. Solid dots in band
  structures indicate the projected $s$ bands. }
\end{figure}

Similar to most of the semiconductors with anti-fluorite or
zinc-blende structures, the low energy electronic properties of
Cd$_3$As$_2$ are mostly determined by the Cd-$5s$ states (conduction
bands) and the As-$4p$ states (valence bands), as shown in
Fig.~\ref{ldaband}. However, there are two distinct features: (1) band-inversion
around $\Gamma$ point with the $s$-state (red solid cycle) lower than
the $p$-states, which is an important sign of non-trivial topology;
(2) semi-metallic with band crossing along the $\Gamma$-Z
direction. This band-crossing is unavoidable, because the two bands
belong to different ($\Lambda_6$ and $\Lambda_7$) representations
respectively, as distinguished by $C_{4}$ rotational symmetry around
$k_z$ axis. The different representation prohibits hybridization
between them, resulting in the protected band-crossing. Furthermore,
the crossing points should locate exactly at the Fermi level due to
charge neutrality requirement, resulting in a 3D Dirac semimetal state
with Fermi surface consisting of a single pair of Fermi points (two
symmetric points along $k_z$ related by TR). Both the structure I and
II share the above common features but with one important difference:
structure I has inversion center but structure II doesn't.

As had been attempted with perturbation method~\cite{vacant}, vacancy
ordering and BZ folding play important role for the band-inversion, in
contrast to the cases of HgTe or Ag$_2$Te~\cite{Ag2Te}, where it was
driven by the the shallow $d$ states. To prove this, we performed
calculations for Cd$_3$As$_2$ in hypothetic anti-fluorite structure
without vacancy (using VCA) and in cubic anti-fluorite structure with
one Cd vacancy, keeping the same lattice constant. We found that the
former (later) has normal (inverted) band ordering at $\Gamma$. At the
BZ boundary X point of hypothetic anti-fluorite structure without
vacancy, there exist shallow $s$ and $p$ states, which are folded onto
the $\Gamma$ point of real structure. Therefore, the hybridization
among the states with the same representation will push the states
away from each other, i.e, make the lowest $s$-state further lower and
highest $p$-state higher, resulting in the band inversion at
$\Gamma$. The band-inversion calculated from GGA is about 0.7 eV for
both Structure I and II. Considering the possible underestimation of
GGA for the $s$-$p$ gap, we have improved the calculations by HSE
method~\cite{hse}, and still found the band-inversion around 0.3 eV,
being consistent with most of the existing experimental evidence, such
as the optical and transport measurements~\cite{invert_exp}.

The calculated band structures of Cd$_3$As$_2$ can be well fitted by
using our modified 8-band Kane model presented in previous section,
and the obtained parameters are listed in Table.I.

\begin{table}[!Ht]
\begin{center}

\begin{tabular}{|c|c|c|c|c| }
  \hline
 $E_s\ (eV)$ & $E_p\ (eV)$  & $\delta\ (eV)$ & $d\ (eV)$ & $\ \Delta\ (eV)$  \\
  \hline
  -0.610367&-0.069191&0.072439 &0.027 & 0.16\\
  \hline
  \hline
  $P\ (eV\AA) $ & $A'\ (eV\AA^2) $ & $L\ (eV\AA^2) $ & $M\ ( eV\AA^2) $ & $N\ ( eV\AA^2) $   \\
  \hline
 6.302242 & 8.013873 & -5.675600  &-7.957689  & -10.757965  \\
  \hline
 \end{tabular}\label{tab:chern111}
\end{center}
\caption{\label{paramts}The fitted parameters for the 8-band model.}
\end{table}

\subsection{Minimal Effective Hamiltonian for the 3D Dirac fermion}

The atomic Cd-$5s$ and As-$4p$ states with SOC can be written as the
states with definite angular momentum $J$ and $J_z$, i. e.,
$|S_{J=\frac{1}{2}},J_z=\pm\frac{1}{2}\rangle$,
$|P_{\frac{3}{2}},\pm\frac{3}{2}\rangle$,
$|P_{\frac{3}{2}},\pm\frac{1}{2}\rangle$,
$|P_{\frac{1}{2}},\pm\frac{1}{2}\rangle$.  In the tetragonal crystal
symmetry, however, the total angular momentum $J$ is no longer a good
quantum number, and the valence $p$-states have complete splitting at
$\Gamma$. The heavy-hole $p$-state
$\Gamma_7=|P_{\frac{3}{2}},\pm\frac{3}{2}\rangle$ and the conduction
$s$-state $\Gamma_6=|S_{\frac{1}{2}},\pm\frac{1}{2}\rangle$ remain to
be the eigenstates at $\Gamma$, while the light-hole states
$|P_{\frac{3}{2}},\pm\frac{1}{2}\rangle$ can mix with the split-off
state $|P_{\frac{1}{2}},\pm\frac{1}{2}\rangle$, forming two new
eigenstates which are irrelevant to the low energy physics here.

\begin{figure}[tbp]
  \includegraphics[clip,scale=0.45]{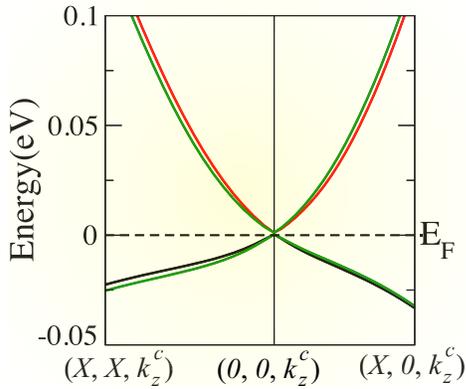}
  \caption{\label{diracband}(Color online) Band dispersions and band-splitting
    in the plane passing through Dirac point (0,0,$k_z^c$) and perpendicular
    to $\Gamma$-Z for structure II. The k-points are indicated in cartesian coordinates.
    $X$ and $k_z^{c}$ are around 0.1 and 0.032 \AA$^{-1}$, respectively.}
\end{figure}

The band inversion nature of Cd$_3$As$_2$ around the $\Gamma$ can be
caught by considering only the minimal basis set of
$|S_\frac{1}{2},\frac{1}{2}\rangle$,
$|P_{\frac{3}{2}},\frac{3}{2}\rangle$,
$|S_\frac{1}{2},-\frac{1}{2}\rangle$ and
$|P_{\frac{3}{2}},-\frac{3}{2}\rangle$ states.  To describe the 3D
Dirac fermion, an effective low energy Hamiltonian $H_{\Gamma}({\vec
  k})$ can therefore be obtained by downfolding the 8-band model into
the subspace spanned by the 4 mimimal basis.  The resulting
$H_\Gamma({\vec k})$ reads,
\begin{eqnarray*}
  H_{\Gamma}({\vec{k}})&=&\epsilon_0(\vec{k}) +\left(\begin{array}{cccc}
      M(\vec{k}) & Ak_{+} & Dk_- & B^{*}(\vec{k}) \\
      Ak_{-} & -M(\vec{k}) &B^{*}(\vec{k})  & 0 \\
      Dk_+ &B(\vec{k})  & M(\vec{k}) & -Ak_{-}\\
      B(\vec{k}) & 0 & -Ak_{+} & -M(\vec{k})
\end{array}\right)
\end{eqnarray*}
where $\epsilon_0({\vec{k}})=C_{0}+C_{1}k_z^2+C_{2}(k_x^{2}+k_y^2)$,
$k_{\pm}=k_{x}\pm ik_{y}$, and
$M(\vec{k})=M_{0}-M_{1}k_z^2-M_{2}(k_{x}^{2}+k_y^{2})$ with parameters
$M_0$, $M_1$, $M_2$ $<$0 to reproduce band inversion.  Under the
tetragonal symmetry, the leading-order term of $B(\vec{k})$ has to
take the high-order form of $(\alpha k_z+\beta D)k_{+}^2$, which can
be neglected if we only consider the expansion up to
$\mathcal{O}(k^2)$.  The terms containing $D$ describe the breaking of
inversion symmetry, which should be zero for structure I. In such
case, the energy dispersion is
$E(\vec{k})=\epsilon_0(\vec{k})\pm\sqrt{M(\vec{k})^{2}+A^{2}k_+k_-}$,
having a pair of four-fold degenerate zero energy Dirac points at
$\vec{k}^c$=(0, 0, $k_z^c=\pm\sqrt{\frac{M_{0}}{M_{1}}}$). Around the
neighborhood of each Dirac point, we can further expand the
Hamiltonian up to $\mathcal{O}(k^2)$. The resulting Hamiltonian is
nothing but the one for 3D massless Dirac fermions, which has
anisotropic linear dispersion $\Delta E(\vec{k}^c+\delta \vec{k})
\approx \pm |\delta \vec{\tilde{k}}|= \sqrt{ (A \delta k_x)^2 + (A
  \delta k_y)^2 + (2M_1k_z^c \delta k_z)^2 }$ (where $\delta \vec{k}$
is the deviation of momentum $\vec{k}$ from the Dirac point
$\vec{k^c}$). The block diagonal form allows us to decouple the
4$\times$4 matrix into two 2$\times$2 matrices, which are Weyl
fermions with degenerate energy but opposite chirality.  For structure
II, which has no inversion symmetry, the non-vanishing D term will
modify the in-plane velocity from $A$ to $A \pm \frac{1}{2} D$, while
keep the Weyl nodes degenerate (because the $C_{4v}$ symmetry along
the Gamma-Z axis contains only 2D irreducible representations for its
double space group). The resulting Dirac semimetal state has 4-fold
degenerate Dirac points, but with splitting of in-plane band
dispersions away from Dirac points (Fig.~\ref{diracband}). This is a
new type of 3D Dirac semimetal state, in contrast to other
examples~\cite{BiO2,Na3Bi}.

A stable Weyl semimetal state with two Weyl nodes separated in
momentum space can be introduced either by lowering the crystal
symmetry from $C_{4v}$ to $C_{4}$, or by breaking the TR
symmetry~\cite{Na3Bi}. In particular, due to the large $g$-factor
(around 30$\sim$40) of Cd$_3$As$_2$~\cite{g-factor}, an
exchange-splitting of $\sim$ 2meV can be introduced by moderate 1T
magnetic field (if we neglect the orbital effects). On the other hand,
if the 4-fold rotational symmetry is broken, a linear leading order
term of $B(\vec{k}) \approx B_1k_z$, will be introduced in the
effective Hamiltonian $H_\Gamma$. In such case, two Weyl nodes will be
coupled together, resulting in a massive Dirac fermions with gap
opening. We have checked the topological invariant $Z_2$ number for
this resulting insulating state, and found it is odd, confirming that
it is a 3D topological insulator~\cite{TIreview,TIreview-2} (due to
the inverted band structure).

\subsection{Surface states and quantum transport properties}

The non-trivial topology and the 3D Dirac cones in Cd$_3$As$_2$
suggest the presence of non-trivial surface states. For this purpose,
we transform the eight-band model into a tight-binding model on a
tetragonal lattice by introducing the substitutions:
 \begin{eqnarray*}
 k_{i}\rightarrow \frac{1}{L_{i}}sin(k_{i}L_{i}), \\
 k_{i}^2\rightarrow\frac{2}{L_{i}^2}(1-cos(k_{i}L_{i})) \\
 (L_{x,y}=a,L_{z}=c).
 \end{eqnarray*}
 Here, $k_i$ refers to $k_x$, $k_y$ and $k_z$, while $a$ and $c$ are
 the tetragonal lattice constants, which are taken as 3.0~\AA\ and
 5.0~\AA, respectively.  This approximation is valid in the vicinity
 of the $\Gamma$ point.  We use an iterative method to obtain the
 surface Green's function of the semi-infinite
 system~\cite{zhangw}. The imaginary part of the surface Green's
 function is the local density of states (LDOS) at the surface.  Since
 GGA underestimate the $s$-$p$ band gap by about 0.4~eV as discussed
 above, we have artificially lifted the on-site energy of $s$-state
 (as given in Table.~\ref{paramts}) by 0.4~eV in the surface state
 calculations.

\begin{figure}[tbp]
\includegraphics[clip,scale=0.6,angle=0]{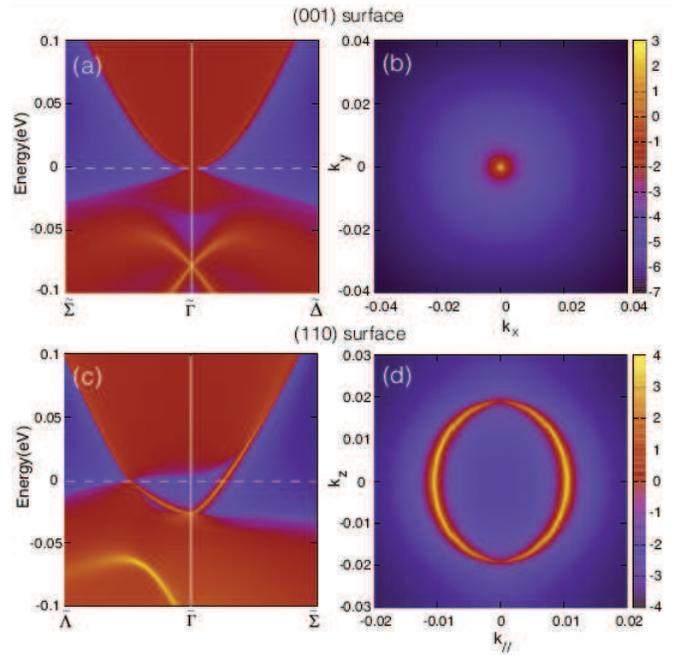}
\caption{\label{surfband}(Color online) The calculated surface states (left panels)
  and corresponding fermi surface (right panels) of Structure II
  Cd$_3$As$_2$ for its (001) (upper panels) and (110) (lower panels)
  surface.}
\end{figure}

 The obtained LDOS on semi-infinite (001) and (110) surfaces of the
 structure II are presented in Fig.~\ref{surfband}.  For (001)
 surface, the surface projection of continuous bulk states superposes
 the non-trivial surface states, and its Fermi surface is just a point
 (Fig.~\ref{surfband}(b)). For (110) surface, however, the non-trivial
 surface states are clearly visible. Its Fermi surface is composed of
 two half-circle Fermi arcs, touching at the singularity points
 ($k_{\parallel}=0$, $\pm k_z^c$) where the surface projection of bulk
 Dirac points exist.

\begin{figure}[tbp]
\includegraphics[clip,scale=0.5]{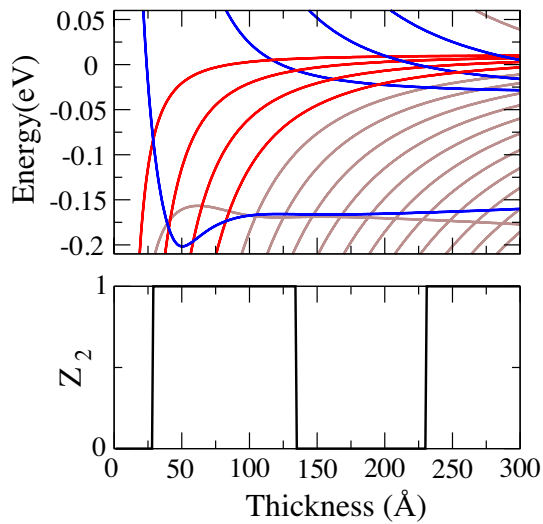}
\caption{\label{wellband}(Color online) The thickness dependence of (a) sub-band
  energies and (b) Z$_2$ number of $z$-oriented quantum well of
  Cd$_3$As$_2$.}
\end{figure}

The existence of 3D Dirac cones in the bulk also implies that we can
expect the QSH effect if we reduce the dimensionality and form the
quantum well structure of Cd$_3$As$_2$. This provides an alternative
compound choice to the existing experiments~\cite{expHgTe,typeII},
which all require extreme conditions up to now. The band structure of
$z$-oriented Cd$_3$As$_2$ thin film of different thickness can be
understood as the bulk bands in different $k_{z}$-fixed planes. The 2D
Z$_2$ number can be nonzero only for limited regions where the band
inversion happens when going from $\Gamma$ to Z. In the quantum-well
structure, those low energy states around $\Gamma$ should be further
quantized into subbands, whose energy levels change as a function of
film thickness (Fig.~\ref{wellband}(a)). When the thickness of the
film is thin enough, the band inversion in the bulk band structure
will be removed entirely by the finite size effect. With the increment
of the film thickness, finite size effect is getting weaker and the
band inversion among these subbands restores subsequently, which leads
to jumps in the Z$_2$ number. Then, depending on the number of band
inversions associated with the sub-bands, the system should cross over
between trivial and nontrivial 2D insulators oscillatorily as a
function of thickness~\cite{oscillatory}
(Fig.~\ref{wellband}(b)). Comparing with HgTe/CdTe quantum
well~\cite{expHgTe}, the first critical thickness of Cd$_3$As$_2$ is
much thinner (2.9nm vs. 6.3nm), suggesting possibly larger gap. To be
a concrete example, 5nm thick Cd$_3$As$_2$ film is a good QSH
insulator with gap more than 100meV. Given the known high mobility of
Cd$_3$As$_2$, it is therefore a good candidate for the QSH
measurement.

Finally we propose that crystalline Cd$_3$As$_2$ with 3D Dirac cone is
an ideal system to check Abrikosov's proposal of quantum MR~\cite{MR},
and may support sizable linear MR even up to room temperature. Quantum
effects become noticeable when the individual Landau levels are
distinct: $\hbar \omega_c$$>$$k_B T$. Dirac system has linear energy
spectrum, and its cyclotron frequency $\omega_c$ should follow the
square root rule ($\omega_c$=$v\sqrt{eB/c}$), in contrast with
non-relativistic system where $\omega_c$ is linear in field $B$. Using
calculated velocity $v$=$3\times 10^7$cm/s for Cd$_3$As$_2$, this then
leads to $\hbar \omega_c\sim$280K for $B$=10T. The linear quantum MR
can be estimated~\cite{MR} as $\Delta\rho(B)/\rho(0)$=$N_iB/\pi n^2
ec\rho(0)$=$N_iB\mu/\pi nc$ (where $\rho(0)$=1/$ne\mu$ is used, $n$ is
carrier density, $\mu$ is mobility, and $N_i$ is the density of
scattering centers). Taking the experimental values
$\mu\sim$1.5m$^2$V$^{-1}$s$^{-1}$ at room temperature, and assuming
$n$ and $N_i$ are in the same order, MR ratio can reach 50\% per 1T
field. If the scattering is not phonon-mediated, we expect that $N_i$
is not sensitive to temperature, then MR is mostly determined by $n$
and $\mu$, which should lead to even enhanced MR at lower temperature.

\section{Conclusion} \label{Conclusion}

In summary, based on the first-principles calculations and effective
model analysis, we have shown that the known compound Cd$_3$As$_2$ is
a symmetry-protected topological semimetal with a pair of 3D Dirac
points (located at the Fermi level) in the bulk and Fermi arcs on the
surfaces. It can be driven into various topologically distinct phases,
such as topological insulator and Weyl semimetal state by symmetry
breakings. In addition, due to its unique 3D Dirac cone type
electronic structure, we can expect the QSH effect in its quantum well
structure and the sizable linear quantum MR even up to the room
temperature. It will be of particular interest, in such a 3D
topological semimetal, to see whether the superconductivity can be
achived by carrier doping or not, because such superconducting state
if obtained may be related to the topological
superconductivity~\cite{TSC}.

\begin{acknowledgments}
  The first-principles calculations are performed by using the
  computers in the Nation Supercomputer Center in Tianjin of China. We
  acknowledge the supports from NSF of China, the 973 program of China
  (No. 2011CBA00108 and 2013CBP21700).
\end{acknowledgments}

\end{document}